\documentclass[a4paper,11pt]{article}
\pdfoutput=1 % if your are submitting a pdflatex (i.e. if you have
\usepackage{jcappub} % for details on the use of the package, please
\usepackage[T1]{fontenc} % if needed

\usepackage{mathtools}
\usepackage{tensor}
\usepackage{amsmath, amsfonts,bm,amssymb}
\usepackage{derivative}
\usepackage{bm}

\usepackage{suffix}

\usepackage{amsmath, amsfonts,bm}
\usepackage{color}
\usepackage{graphicx}
\usepackage[usenames,dvipsnames]{xcolor}
\usepackage{subfig}
\definecolor{myred}{rgb}{0.9, 0.17, 0.31}
\usepackage{geometry}
\usepackage[dvipsnames]{xcolor}
\usepackage{hyperref}% add hypertext capabilities
\hypersetup{
    colorlinks=true,
    allcolors=RedViolet
    }

\newcommand\mathcomma{\,,}
\newcommand\mathperiod{\,.}

\DeclareMathAlphabet{\mathup}{OT1}{\familydefault}{m}{n}

\DeclareMathOperator{\sech}{sech}

\def\dd{\mathrm{d}}

\newcommand{\be}{\begin{equation}} 
\newcommand{\ee}{\end{equation}}

\title{The dynamics of three-forms in thick branes}

\author[a,1]{Jake E.B. Gordin\note{Corresponding author.},}
\author[b]{Kelly MacDevette,}
\author[b]{Jenna Bruton,}

\affiliation[a]{Department of Physics, \\University of Oslo, \\ Box 1048, N-0316 Oslo, Norway}
\affiliation[b]{Cosmology and Gravity Group, \\Department of Mathematics and Applied Mathematics,\\
University of Cape Town, \\Rondebosch 7700, Cape Town, South Africa}

\emailAdd{j.e.b.gordin@fys.uio.no}
\emailAdd{mcdkel004@myuct.ac.za}
\emailAdd{brtjen003@myuct.ac.za}
\abstract{In this work, we investigate thick brane models with a single three-form field. We find novel solutions for thick braneworlds where only three-forms exist and interact gravitationally in the bulk, both with and without matter fields. We use an additional scalar field as proxy for the matter fields. As an initial study, we consider the results here in contrast to the single scalar field thick braneworld case. The properties of the specific three-form parameterisation limits the freedom we have to choose the form of the warp factor, leading to a closed system of equations with nontrivial yet unstable solutions. The
stability of the gravitational sector for thick brane three-forms is investigated and the models are shown to be unstable against
small perturbations of the metric, further indicating that three-forms cannot exist stably in thick braneworld settings.}

\begin{document}
\maketitle
\flushbottom
\newpage
\section{Introduction}
\label{sec:intro}
%%%%%%%%%%%%%%%%%%%%%%%%%%%%%%%%%%%%%%%%%%%%%%%%%%%%%%%%%%%%%%%%%%%%%%

One of the perennial problems in cosmology is the question ``What provided the initial seeds for structure formation?'' The theory of cosmological perturbations (for a thorough reference, see \cite{Peter:1208401}) aims to describe the evolution of the universe \textit{given} some collection of structures, but it does not account for the cause of the primordial density fluctuations. Inflation, coupled with quantum fluctuations (see \cite{Baumann:2009ds}), is currently the most promising contender. Inflation is attractive for its simplicity -- we only need to add in a scalar field -- and its ability to directly solve open theoretical issues with the big bang model. A scalar field is not the only possibility as the dynamical object driving inflation. Given that the empirical status of inflation is at present inconclusive (see \cite{achúcarro2022inflation} for an overview of the observational status and prospects), there is considerable theoretical degeneracy among possible models. Scalars are favoured for reasons of parsimony, but there are no experimental reasons to discount alternatives.\\

In recent years, one such alternative to be considered is three-forms (for a subset of work on three-form inflation, see \cite{Barros:2015evi,Mulryne:2012ax,Koivisto:2009fb,SravanKumar:2016biw,Kumar:2014oka,Koivisto:2009sd,Germani:2009iq,DeFelice:2012jt}). Moreover, much like how scalar fields are also dark energy candidates, so too are three-forms \cite{Koivisto:2012xm,Koivisto:2009fb}. Three-forms have also been studied in other contexts, namely spherical objects (black holes, stars, and wormholes \cite{Barros:2020ghz,Barros:2018lca,Barros:2021jbt,Bouhmadi-Lopez:2020wve,Bouhmadi-Lopez:2021zwt}), singularities in general relativity \cite{Bouhmadi-Lopez:2018lly,Morais:2017vlf}, for use in screened cosmologies \cite{Barreiro:2016aln}, and for studies on alternatives to FLRW spacetimes \cite{Normann:2017aav}. More infomation about the general three-form formalism can be found in \cite{Koivisto:2009ew,BeltranAlmeida:2018nin,Wongjun:2016tva}.\\

A three-form is a three-indexed tensor field, $A_{ABC}$. Their inclusion as an alternative field is not without precedent, as they occur naturally in supersymmetry and string theory models (an overview can be found in \cite{Farakos_2018}.) They are not mere mathematical oddities: they produce distinct signatures for all the aforementioned effects (inflation, dark energy, etc.) Although non-scalar inflationary mechanisms have been considered (for example, vector inflation \cite{Koivisto_2008}), we choose to pay three-forms particular attention for the following reasons: (a) vectors and other two-forms fields (e.g. Kalb-Ramond field) have been more considerably studied than three-forms (although not as thoroughly as scalars)
%much as scalars, but still)%
and (b) forms of four or higher can be shown to be equivalent to scalar fields \cite{gupta2011dark}. Given their origin in string theoretic models, it is natural to study their dynamics in dimensions higher than 4. We could perform the analysis in the full string theoretic 11 dimensions, but it is useful to have some means of comparison to scalars. For that reason, we will study three-forms in branes (i.e. 5D). \\

Several studies have been done already in branes, typically involving additional degrees of freedom in the gravitational theory (like scalar fields) \cite{Bazeia_2006, Afonso_2006, Hendi:2020qkk}  or some modification of the theory itself ($f(R)$, conformal gravity, see \cite{Rosa:2021tei} and references in their introduction.) An overview of scalars and thick brane solutions can be found in \cite{Dzhunushaliev_2010}. Few studies, however, have been done on three-forms in branes. There is an analysis of three-form cosmological solutions in the Randall-Sundrum II braneworld scenario \cite{Barros:2015evi}. This is an example of a thin-brane model \cite{Dzhunushaliev_2010}. To the authors' knowledge, there is no study of three-forms in thick braneworlds. In this paper we aim to plug that gap. \\

As the beginnings of a preliminary analysis of three-forms in branes, we lay the groundwork: we will construct the model, that of a three-form in a warped flat thick brane, find the three-form analogue of the Klein-Gordon (3KG) equation of motion, the Einstein field equations (EFEs), analyse their solutions and their stability against metric perturbations, and do so for both a matter-free and matter-full setup. Our analysis will reveal crucial differences between three-forms and scalars in thick branes: extra terms in the equations of motion, degeneracy in the EFEs, and most saliently, instabilities. The classical thick brane background is not stable against linear perturbations when three-forms inhabit the bulk -- this is in contrast to most scalar field models. Similarly, the overall system of equations governing three-form dynamics are not under-determined, like scalars. They must be solved directly as is, subject only to a choice of three-form dual parameterisation, and no freedom to fix the warp factor. We will see that the solutions reflect the instabilities found in the perturbative analysis. It appears as though, barring some other modification, or high degree of fine-tuning, three-forms are not stable in thick braneworlds. There will be one caveat to this whole discussion: the choice of three-form dual parameterisation. \\

We will elaborate on the above and on the possible implications in the remainder of the paper. In Section \ref{sec:model} we construct the model, writing down the relevant equations. In Section \ref{sec:pert} we analyse the background stability via perturbations. We solve the dynamical system in Section \ref{sec:sols}. We discuss the results in Section \ref{sec:discussion}. We supplement the stability discussion in the appendix, via an application of the formalism developed in \cite{Hendi:2020qkk} for a general $q$-form.\\

%%%%%%%%%%%%%%%%%%%%%%%%%%%%%%%%%%%%%%%%%%%%%%%%%%%%%%%%%%%%%%%%%%%%%%
\section{Thick brane model for three-forms}
\label{sec:model}

Throughout this work we use notation where (capital) Latin indices run through 5D, i.e. ${B\in\{0,1,2,3,4\}}$ and Greek indices run the standard 4D coordinates, i.e. ${\mu\in\{0,1,2,3\}}$. The square of a tensor denotes contraction of all the indices, i.e. ${A^2=A_{ABC}A^{ABC}}$, and a circle denotes contraction of all but the first index, i.e. ${\left( A\circ A\right)_{AB}=A_A{}^{CD}A_{BCD}}$. We work in units such that $c = 1$. Additionally, we will eschew the usual convention in braneworld papers and use $\mathcal{W}$ to denote the warp function, as the three-form potential is typically denoted by $\textbf{A}$. Boldface quantities refer to 5D vectors.

\subsection{Three-form action}

We consider the following action for a three-form field $A_{ABC}$  minimally coupled to Einstein gravity in a 5D spacetime
\be\label{action}
S=\int d^4xdy\sqrt{-g}\left[ \frac{1}{2\kappa^2}R-\frac{1}{48}F^2-V(A^2) \right]\mathcomma
\ee
where $y$ identifies the coordinate of the fifth dimension (or `bulk'), ${g=\det g_{AB}}$ is the determinant of the metric, $R$ is the standard curvature scalar, ${\kappa^2=8\pi G_5}$ with $G_5$ being the 5D Newton's constant, $V(A^2)$ is the three-form self interacting potential and ${\bf F=\dd A}$ is the strength tensor of the three-form, with components,
\be
F_{ABCD} = 4\nabla_{[A}A_{BCD]}\mathperiod
\ee
In a torsion-free spacetime the covariant derivatives in the above equation can be replaced with ordinary partial ones, due to the symmetric nature of the connection \cite{Wald:106274}. Here, the strength tensor plays the same role for the three-form theory as the kinetic field strength ${\partial_{\mu}\phi}$ does for standard scalar field theory or as $F_{\mu\nu}$ for classical Maxwell's electromagnetism, corresponding to a zero ($\phi$) and one-form ($A_{\mu}$) respectively. This three-form will naturally endow its field $A_{ABC}$ with dynamics, depending on the choice of metric $g_{AB}$.

\subsection{Metric and equations of motion}

In this paper we will consider a conformally flat \cite{Cs_ki_2000} 5D brane spacetime, with line element,
\be\label{metric}
ds^2 = e^{2\mathcal{W}(y)}\eta_{\mu\nu}dx^{\mu}dx^{\nu} + dy^2\mathcomma
\ee
where $\eta_{\mu\nu}$ is the Minkowski 4D metric, with signature ${(-,+,+,+)}$, and the prefactor $e^{2\mathcal{W}(y)}$ is the so-called ``warp factor" with $\mathcal{W}(y)$ being the ``warp function". \\
%%%%%%%%%%%%%%%%%%%%%%%%%%%%%%%%%%%%%%%%%%%%%%%%%%%%%%%%%%%%%%%%%%%%%%
%\subsection{The three-form in five dimensions} 
In 5D the three-form dual, ${ \star A_{ABC} =B_{AB}}$, is a two-form, $B_{AB}$, with components
\be\label{dual}
 B^{AB}=(\star A)^{AB} = \frac{1}{3!}\frac{1}{\sqrt{-g}}\epsilon^{ABCDE}A_{CDE}\mathcomma
\ee
where here $\epsilon$ denotes the 5D Levi-Civita symbol \cite{Barros:2015evi}. We may now introduce a scalar function ${\chi (y)}$ that parametrizes $B_{AB}$ and depends only on the fifth dimension, $y$. Using eq. \eqref{metric}, the dual \eqref{dual} has the following non-vanishing components:
\be\label{dualansatz}
B_{0y}=-B_{y0}=e^{\mathcal{W}(y)}\chi (y)\mathperiod
\ee

\textit{This is a fixed, free choice.} This antisymmetric ansatz for the dual vector greatly simplifies the equations and allows the three-form components to be completely determined by the field $\chi (y)$. Inverting ~\eqref{dual} gives
\be\label{3form_comp}
A_{ABC} = \sqrt{-g}\,\epsilon_{ABCDE}B^{DE}\mathcomma
\ee
which has the non-zero components:
\begin{gather}\label{3form_comp_explicit}
A_{123}=A_{231}=A_{312}=-A_{132}=-A_{321}=-A_{213}\nonumber\\
=2e^{3\mathcal{W}}\chi\mathperiod
\end{gather}
Finally, we can calculate the invariants:
\begin{eqnarray}
A^2 &=& 24\chi^2\mathcomma\label{eq:invariantsA}\\
F^2 &=& 96\left( \chi'+3\mathcal{W}'\chi \right)^2\mathcomma
\end{eqnarray}
where a prime denotes a derivative with respect to $y$, i.e. ${\chi' = d\chi/dy}$.\\  
    
With the three-form invariants \eqref{eq:invariantsA} and the brane metric \eqref{metric} we are now ready to calculate the equations of motion using eq. \eqref{action}. Varying the total action ~\eqref{action} with respect to the three-form yields the following equations of motion,
% \be\label{motion_A}
% \nabla\cdot {\bf F} = 12\frac{dV}{dA^2}{\bf A}\mathcomma
% \ee
% or, in component form,
\be\label{motion_A}
\nabla_A F^A{}_{BCD} = 12\frac{dV}{dA^2}A_{BCD}\mathperiod
\ee
Due to the antisymmetric nature of the ${\bf A}$ field, ${\bf F}$ is a closed differential form, i.e. ${\bf dF}=0$. \\

% \subsection{Components of system of equations}

Plugging the metric \eqref{metric} into eq.~\eqref{motion_A} one may express the equations of motion in terms of the $\chi$ field as
\be\label{eom}
\chi ''+ 4 \mathcal{W}' \chi '+3 \chi  \left(\mathcal{W}''+\mathcal{W}'^2\right)-\frac{1}{4}\frac{dV}{d\chi} = 0\mathcomma
\ee
where ${V_{\chi}=dV/d\chi}$. Note that is in essence the equation of motion for the three-form, via the parameterisation $B_{AB}$ viz. its dual, $A_{ABC}$. The Einstein field equations can be computed by the variation of eq.~\eqref{action} with respect to the metric $g^{AB}$. They are
\be\label{field_eqs}
G_{AB}=\kappa^2T_{AB}\mathcomma
\ee
where $G_{AB}$ is the standard Einstein tensor and the stress-energy tensor is sourced entirely by the three-form,
\be\label{3f_EM}
T_{AB} = \frac{1}{6}\left( F\circ F\right)_{AB}+6\frac{dV}{dA^2}\left( A\circ A\right)_{AB}+g_{AB}\mathcal{L}_{3f}\mathcomma
\ee
where from eq.~\eqref{action} we identify the three-form Lagrangian density as,
\be
\mathcal{L}_{3f}=-\frac{1}{48}F^2-V(A^2)\mathperiod
\ee
With our brane metric \eqref{metric} the components of $T_{AB}$ are:
\begin{eqnarray}
    T^0{}_0 &=& -V-\frac{1}{48}F^2\mathcomma \\
    T^1{}_1 &=& T^2{}_2=T^3{}_3=-T^0{}_0-2V+\chi V_{\chi}\mathcomma\quad \\
    T^y{}_y &=& -T^0_0-2V\mathcomma
\end{eqnarray}

with trace:
\be
T=g^{AB}T_{AB} = 6\left(\chi' + 3\mathcal{W}'\chi\right)^2 +3\chi V_{\chi}-5V\mathcomma
\ee
and the components of the Einstein tensor:
\begin{eqnarray}\label{einsteintensor}
    G^0{}_0 &=&  3\left(\mathcal{W}''+2\mathcal{W}'^2\right)\mathcomma \\
    G^1{}_1&=&G^2{}_2=G^3{}_3=G^0{}_0\mathcomma\\
    G^y{}_y &=& 6 \mathcal{W}'^2.
\end{eqnarray}

We will now set $\kappa^2 =1$ (cf. \cite{Rosa:2021tei} or \cite{Afonso_2006}, who set it to $2$). Our full system of equations, determining the dynamics of both the metric and three-form, is 

\be\label{eomf}
\chi ''+ 4 \mathcal{W}' \chi '+3 \chi  \left(\mathcal{W}''+\mathcal{W}'^2\right)-\frac{1}{4}\frac{dV}{d\chi} = 0\mathcomma
\ee

\be\label{efef}
\begin{split}
    3\left(\mathcal{W}''+2\mathcal{W}'^2\right) &= -V-\frac{1}{48}F^2\mathcomma\\
    3\left(\mathcal{W}''+2\mathcal{W}'^2\right) & =  \frac{1}{48}F^2 -V+\chi V_{\chi}\mathcomma\\
    6 \mathcal{W}'^2 &= \frac{1}{48}F^2 - V.
\end{split}
\ee

\subsection{Comparison with the scalar field case}

It is interesting, on a purely superficial level, to compare our system of equations \eqref{eomf}, \eqref{efef} to the scalar field case. For a minimally coupled scalar field $\varphi$ in our metric, the full system reads \cite{Afonso_2006, Peyravi_2016}

\begin{equation}\label{scalareomf}
\begin{aligned}
3 \mathcal{W}^{\prime \prime}+6{\mathcal{W}^{\prime}}^2 & =-\left[\frac{1}{2} \varphi^{\prime 2}+V(\varphi)\right], \\
6{\mathcal{W}^{\prime}}^2 & =\left[\frac{1}{2} \varphi^{\prime 2}-V(\varphi)\right], \\
\varphi^{\prime \prime}+4 \mathcal{W}^{\prime} \varphi^{\prime} & =\frac{d V(\varphi)}{d \varphi}
\end{aligned}
\end{equation}

The Einstein tensor is obviously the same, but the stress-energy tensor of the scalar field has better degrees of symmetry: its 4D components are identical, so there is only one EFE. The $00$ stress-energy component is the same as the three-form: it's simply the Lagrangian of the field in question.\\

The differences between the three-form and scalar occur at the spatial (3D) level: the three-form's stress-energy tensor is quite different to the scalar field one, containing not only an opposite kinetic sign but a potential derivative. Finally, the $yy$ component is the same as the three-form case. The true EFE difference manifests in the spatial 3D behaviour, not in the bulk. This is reflective of the nonzero components for the three-form in our parameterisation, eq. \eqref{3form_comp_explicit}. \\

The equation of motion is also different: there's an additional $3 \chi  \left(\mathcal{W}''+\mathcal{W}'^2\right)$ term and some scaling of the potential derivative for the three-form. This however is not only generically expected for a tensor field as compared to a scalar, but also the form our three-form ``Klein-Gordon'' takes is highly dependant on the choice of parameterisation. This is not the case for the EFEs. This is crucial since the total system for the three-form, unlike the scalar field, is \textit{not} overdetermined. To wit, you can't derive one equation in the system \eqref{eomf}, \eqref{efef} from the other three -- you \textit{do} have this liberty in eq. \eqref{scalareomf}. This fact is exploited in typical scalar field studies to fix a known warp function and obtain closed solutions (\cite{Dzhunushaliev_2010}; see, for example, \cite{Rosa:2021tei}). We can reduce our system to three independent equations in three unknown variables:
%Our system of four equations can be written in terms of three equations. This means we have three equations and three unknowns: 
$\chi, V, \mathcal{W}$. Thus the system is not under-determined and we are not free to impose any further choices.

\section{Stability of the graviton through perturbations \label{sec:pert}}

In braneworld models, one question that must be asked and answered is are the zero-order modes stable against perturbations? That is to say, for a metric perturbation $h_{\mu\nu}$, obeying a Schrodinger-like equation of motion, does the Sturm–Liouville operator admit negative energy states? If so, the classical background is not stable \cite{DeWolfe_2000}. We follow \cite{DeWolfe_2000, Bazeia_2015}. The perturbed metric is only in 4D, the bulk remains unchanged. The metric is therefore 

\begin{equation}
d s^2=e^{2 \mathcal{W}(y)}\left[\eta_{\mu \nu}+h_{\mu \nu}(x, y)\right] d x^\mu d x^\nu+d y^2
\end{equation}

The perturbed EFEs will therefore be 

\begin{equation}
   \delta R_{AB}=\delta{T}_{AB} -\frac{1}{3} \delta g_{AB} \delta T_C^C.
\end{equation}

The perturbed Ricci tensor is the same for any minimally coupled braneworld model (here, we need only the 4D component, since it is this that enables us to analyse stability):

    \begin{equation}\label{perturbedriccitensor}
\begin{aligned}
\delta R_{\mu\nu}= & e^{2 \mathcal{W}}\left(\frac{1}{2} \partial_y^2+2  \mathcal{W}^{\prime} \partial_y+ \mathcal{W}^{\prime \prime}+4  \mathcal{W}^{\prime 2}\right) h_{\mu\nu}+\frac{1}{2} \eta_{\mu\nu} e^{2  \mathcal{W}}  \mathcal{W}^{\prime} \partial_y\left(\eta^{\alpha\beta} h_{\alpha\beta}\right)-\frac{1}{2} \square h_{\mu\nu} \\
& -\frac{1}{2} \eta^{
\alpha\beta
}\left(\partial_\mu \partial_\nu h_{\alpha\beta}-\partial_\mu \partial_\alpha h_{\nu\beta}-\partial_\nu \partial_\alpha h_{\mu \beta}\right)
\end{aligned}
\end{equation}

where $\square$ is w.r.t to $g_{\mu\nu}$, not $g_{AB}$. The transverse-traceless gauge \cite{carroll2003spacetime} eliminates the second and fourth term, leaving 

\begin{equation}
\begin{split}
    e^{2 \mathcal{W}}\left(\frac{1}{2} \partial_y^2+2 \mathcal{W}^{\prime} \partial_y+\mathcal{W}^{\prime \prime}+4 \mathcal{W}^{\prime 2}\right) h_{\mu\nu}-\frac{1}{2} \square h_{\mu\nu}  = \delta{T}_{\mu\nu} -\frac{1}{3} \delta g_{\mu\nu} \delta T_C^C \mathcomma
\end{split}    
\end{equation}

where the trace is with respect to all indices. %since you can't have a separated out trace %\jg{right..?}.
The key calculation is the RHS. For the scalar field case, we are able to eliminate $\mathcal{W}'' + 4 \mathcal{W}'$, giving us the following wave equation for $h_{\mu\nu}$ \cite{Rosa:2021tei}

\begin{equation}
\left[-\frac{d^2}{d z^2}+u(z)\right] \bar{h}_{\mu \nu}(z)=p^2 \bar{h}_{\mu \nu}(z)
\end{equation}

for 

\begin{equation}
    {h}_{\mu\nu} = e^{-i p \cdot x} e^{-3 \mathcal{W}(z) / 2}  \bar{h}_{\mu \nu}(z)
\end{equation}

This operator $-\partial^2_z + u(z)$ is bounded from below, and the lowest mode $p^2 =0$ is normalisable. The crucial step is being able to eliminate $\mathcal{W}'' + 4 \mathcal{W}'$ in eq. \eqref{perturbedriccitensor}. If this is not possible, the differential operator changes. We now return to the RHS for our three-form $S_{\mu\nu} =  \delta{T}_{\mu\nu} -\frac{1}{3} \delta g_{\mu\nu} \delta T_C^C $, 

\begin{equation}\label{sourceterm}
\begin{split}
    S_{\mu\nu} = \frac{1}{6}\delta \left( F\circ F\right)_{\mu\nu}+6\delta \frac{dV}{dA^2}\left( A\circ A\right)_{\mu\nu} +6 \frac{dV}{dA^2}\delta \left( A\circ A\right)_{\mu\nu}+\eta_{\mu\nu} \delta \mathcal{L}_{3f} \\+e^{2\mathcal{W}}h_{\mu\nu} \delta \mathcal{L}_{3f}^{0} - \frac{1}{3} +e^{2\mathcal{W}}h_{\mu\nu}T
\end{split}
\end{equation}

Explicitly, we note that the nonzero terms are

\begin{equation}
    \left( A\circ A\right)_{\mu\nu} = \left( A\circ A\right)_{ii} = 8 e^{2 \mathcal{W}} \chi^2.
\end{equation}

\begin{equation}
\begin{split}
     \left( F\circ F\right)_{\mu\nu} = \left( F\circ F\right)_{ii} =     \left( F\circ F\right)_{yy} = 
     24 e^{2 \mathcal{W}} \left(3 \chi\mathcal{W}'+\chi '\right)^2 
\end{split}
\end{equation}

where $i = \{1,2,3\}$. The three-form field has vector and scalar perturbations. But in our case we only have (cf. eq. \eqref{3form_comp_explicit}) perturbations \cite{Koivisto_2013}

\begin{equation}
\begin{aligned}
A_{i j k}= \epsilon_{i j k}\left(\chi_0 + \delta \chi\right).
\end{aligned}
\end{equation} \medskip

% We can write these in terms of the standard invariants, 

% \begin{equation}
%     \begin{split}
%         \left( A\circ A\right)_{\mu\nu} &= \frac{1}{4} e^{2 \mathcal{W}}A^2 \quad \text{for $\mu=\nu = 1,2,3$} \\
%  \left( F\circ F\right)_{\mu\nu}&= \frac{1}{4} e^{2 \mathcal{W}}F^2 \quad \text{for $\mu= \nu = 1,2,3,y$}
%     \end{split}
% \end{equation}

As such, the terms $\delta \left( F\circ F\right)_{\mu\nu}$ and $\delta \left( A\circ A\right)_{\mu\nu}$ will contain several perturbed terms, but none proportional to $h_{\mu\nu}$. To linear order then, 

\begin{equation}
    \begin{split}
       \delta \left( A\circ A\right)_{\mu\nu} &= 8 e^{2 \mathcal{W}} (2\chi_0 \delta \chi ) \\
 \delta \left( F\circ F\right)_{\mu\nu}&=  48 e^{2 \mathcal{W}} \left(3  \delta \chi \mathcal{W}'+  \delta \chi' \right)^2
    \end{split}
\end{equation}

% The terms on the first line are all proportional to $\chi_0 + \delta \chi$, but not $h_{\mu\nu}$. The first term on the second line is obviously $\propto h_{\mu\nu}$, as well as the second term if the trace is the zeroth-order trace. 

This is similarly true for $\eta_{\mu\nu} \delta \mathcal{L}_{3f}$, since $\mathcal{L}_{3f}$ contains only $F^2$ and $V$ terms. The terms of concern are therefore the last two in the source term \eqref{sourceterm}, 

\begin{equation}
    S_{\mu\nu} \propto e^{2 \mathcal{W}}h_{\mu\nu} \left(-\frac{1}{48}F^2-V(A^2)\right)^{(0)} - \frac{1}{3} e^{2 \mathcal{W}}h_{\mu\nu} T^{(0)}.
\end{equation}

The term we require, in order to cancel out the term we want, is from eq.~\eqref{efef}

\begin{equation}
    \mathcal{W}'' + 4\mathcal{W}' = \frac{1}{144}F^2 - V \quad %(\text{platonic ideal}).
\end{equation}

Thus, we have, after lots of tedious algebra,

\begin{equation}\label{pertsourceextraterm}
    S_{\mu\nu} \propto  \frac{2}{3}e^{2 \mathcal{W}}h_{\mu\nu} V.
\end{equation}

This does not equal the ideal term, and so of course doesn't cancel. The equation of motion for $h_{\mu\nu}$ is therefore 

\begin{equation}
    \begin{split}
        e^{2 \mathcal{W}}\left(\frac{1}{2} \partial_y^2+2 \mathcal{W}^{\prime} \partial_y+\mathcal{W}^{\prime \prime}+4 \mathcal{W}^{\prime 2} - \frac{2}{3} V\right) h_{\mu\nu}-\frac{1}{2} \square h_{\mu\nu}  = \delta \tilde{T}_{\mu\nu}.
    \end{split}
\end{equation}

% The source term here is perturbed too. Explicitly, we note that 

% \begin{equation}
%     \left( A\circ A\right)_{\mu\nu} = \left( A\circ A\right)_{11} = \left( A\circ A\right)_{22} = \left( A\circ A\right)_{33} = 8 e^{2 \mathcal{W}} \chi^2.
% \end{equation}

% \begin{equation}
% \begin{split}
%      \left( F\circ F\right)_{\mu\nu} = &\left( F\circ F\right)_{11} = \left( F\circ F\right)_{22} = \left( F\circ F\right)_{33} = \\
%      &\left( F\circ F\right)_{yy} = 
%      24 e^{2 \mathcal{W}} \left(3 \chi\mathcal{W}'+\chi '\right)^2 
% \end{split}
% \end{equation}

% We can write these in terms of the standard invariants, 

% \begin{equation}
%     \begin{split}
%         \left( A\circ A\right)_{\mu\nu} &= \frac{1}{4} e^{2 \mathcal{W}}A^2 \quad \text{for $\mu=\nu = 1,2,3$} \\
%  \left( F\circ F\right)_{\mu\nu}&= \frac{1}{4} e^{2 \mathcal{W}}F^2 \quad \text{for $\mu= \nu = 1,2,3,y$}
%     \end{split}
% \end{equation}

% The three-form field will be perturbed here, but as we can see there is no metric dependence. To linear order then, 

% \begin{equation}
%     \begin{split}
%         \left( A\circ A\right)_{\mu\nu} &= \frac{1}{4} e^{2 \mathcal{W}} \left( A_0^2 + \delta A^2\right) \\
%  \left( F\circ F\right)_{\mu\nu}&= \frac{1}{4} e^{2 \mathcal{W}} \left(F_0^2 + \delta F^2\right)
%     \end{split}
% \end{equation}

% We note then that the shorthand $\left( A\circ A\right)_{\mu\nu}$ and $\left( F\circ F\right)_{\mu\nu}$ refers to the total linearly perturbed terms. 

where 

\begin{equation}
    \delta \tilde{T}_{\mu\nu} = \frac{1}{6}\delta \left( F\circ F\right)_{\mu\nu}+6\delta \frac{dV}{dA^2}\left( A\circ A\right)_{\mu\nu} +6 \frac{dV}{dA^2}\delta \left( A\circ A\right)_{\mu\nu}+\eta_{\mu\nu} \delta \mathcal{L}_{3f} 
\end{equation}

Next, rearrangement yields

\begin{equation}
    \begin{split}
       \left( \partial_y^2+4 \mathcal{W}^{\prime} \partial_y+2\mathcal{W}^{\prime \prime}+8 \mathcal{W}^{\prime 2} - \frac{4}{3} V - e^{-2 \mathcal{W}} \square \right) h_{\mu\nu} = 2e^{-2 \mathcal{W}} \delta \tilde{T}_{\mu\nu}
    \end{split}
\end{equation}

% \begin{equation}
%     \begin{split}
%        \left( \partial_y^2+4 \mathcal{W}^{\prime} \partial_y+2\mathcal{W}^{\prime \prime}+8 \mathcal{W}^{\prime 2} - \frac{4}{3} V - e^{-2 \mathcal{W}} \square \right) h_{\mu\nu}\\  = 2e^{-2 \mathcal{W}} \left[\frac{1}{6}\left( F\circ F\right)_{\mu\nu}+6\frac{dV}{dA^2}\left( A\circ A\right)_{\mu\nu}\right]
%     \end{split}
% \end{equation}

This a sourced wave equation, where the particulars of the source term are immaterial. More pertinent to our analysis is the form of the differential operator after the change of variable, $dz = e^{-\mathcal{W}(y)}dy$ and

\begin{equation}\label{perttransform}
    {h}_{\mu\nu} = e^{-i p \cdot x} e^{-3 \mathcal{W}(z) / 2}  \bar{h}_{\mu \nu}(z).
\end{equation}

We can perform this variable transformation step-by-step. Firstly, the d'Alembertian only acts on $e^{-i p \cdot x}$ and, with our metric signature, gives $-p^2$. So, 

\begin{equation}
    \square h_{\mu\nu} = - p^2 e^{-i p \cdot x} e^{-3 \mathcal{W}(z) / 2}  \bar{h}_{\mu \nu}(z) \mathperiod
\end{equation}

Next, our derivatives become \cite{DeWolfe_2000}

\begin{equation}
    \frac{d}{dy} = e^{-2W(z)} \frac{3}{4} \frac{d}{dz}, \quad  \frac{d^2}{dy^2} = e^{-2W(z)} \frac{d^2}{dz^2}
\end{equation}

%\jg{Know how this is done: comes changing $\mathcal{W}' \rightarrow W'$}\\
where now $W$ is a function of z, and a prime is a now a derivative w.r.t to $z$ (the change from $\mathcal{W}(y) \rightarrow W(z)$ is given by integrating the above equation). The exponential can be moved from the RHS, leaving 

\begin{equation}
    \begin{split}
       \left( \partial_z^2+3 W^{\prime} \partial_y+ e^{2W}\left[2W^{\prime \prime}+8 W^{\prime 2} - \frac{4}{3} V\right] + p^2 \right) h_{\mu\nu}  = 2 \delta \tilde{T}_{\mu\nu}
    \end{split}
\end{equation}

Applying the LHS to eq.~\eqref{perttransform} \\

   \begin{equation}
\begin{split}
    3 W'  \partial_z  \left(e^{-i p \cdot x} e^{-3 W(z) / 2}  \bar{h}_{\mu \nu}(z)\right) = e^{-i p \cdot x}  3 W' e^{-3 W(z) / 2} \bar{h}'_{\mu \nu} - e^{-i p \cdot x}  \frac{9}{2} W'^2 e^{- 3W(z) / 2} \bar{h}_{\mu \nu};
\end{split} 
\end{equation}

   \begin{equation}
\begin{split}
      \partial^2_z  \left(e^{-i p \cdot x} e^{-3 W(z) / 2}  \bar{h}_{\mu \nu}(z)\right) = e^{-i p \cdot x} e^{-3 W(z) / 2} \bar{h}''_{\mu \nu} - e^{-i p \cdot x}  3 W' e^{-3 W(z) / 2} \bar{h}'_{\mu \nu} \\+  e^{-i p \cdot x}  \frac{9}{4} W'^2 e^{-3 W(z) / 2} \bar{h}_{\mu \nu} -  e^{-i p \cdot x}  \frac{3}{2} W'' e^{-3 W(z) / 2} \bar{h}_{\mu \nu}
\end{split} 
\end{equation}

We see the $h'_{\mu\nu}$ terms cancel, leaving us to divide through by $e^{-i p \cdot x} e^{-3 W(z) / 2}$:

\begin{equation}\label{gravitoneom}
\begin{split}
       &\left(  \frac{d^2}{dz^2} + u(z) +  p^2 \right) \bar{h}_{\mu \nu}  = \bar{S}_{\mu\nu};\\
       &u(z) = \left[8 e^{2W}-  \frac{9}{4}\right] W'^2 + \left[2  e^{2W}  -   \frac{3}{2}\right]   W''     - e^{2W} \frac{4}{3} V;\\
       &\bar{S}_{\mu\nu}=e^{i p \cdot x} e^{3 W(z) / 2} 2\delta \tilde{T}_{\mu\nu}
\end{split}
\end{equation}

This is now a Schrodinger-type equation. As a sanity-check, when there is no three-form the RHS vanishes, and the effective potential $u(z)$ would match the scalar field case since the extra $e^{2W}$ terms would not appear in earlier steps in the derivation. The three-form has changed the differential operator to, for $p>0$,

\begin{equation}
    D^2 = - \frac{d^2}{dz^2} - u(z).
\end{equation}

This operator is \textit{not} factorisable and so the gravity sector on the brane is not linearly stable \cite{DeWolfe_2000, Rosa:2021tei, Bazeia_2015, Cs_ki_2000}. Even with a source term this conclusion holds: the term $p^2$ corresponds to the energy eigenvalue, and this term is not necessarily bounded from below. There may be regions where this is the case, but the background metric is unstable against perturbations, and so the energy can become unbounded at certain points. With the introduction of the three-form field, the gravity sector of the braneworld is linearly unstable. We shall expect, from this, high sensitivity in the dynamics of the three-form away from very stable fixed points. That is to say, instabilities in growth once there is sufficient kinetic energy to move the three-form along its potential away from such points; and that the required energy needn't be high. We shall confirm this hypothesis in the next section.

% Even with a source term this conclusion holds: the combined term $p^2 + S_{\mu\nu}$ might be positive at some points, but it would need to positive everywhere. If we set $\chi=0$ at a point and kill the RHS, at that point $p^2$ would be known to be negative. \jb{This apparently \textit{is} physics} \jg{Lol, the only bit of physics in this whole thing.} \kelly{Wait I'm not sure about this last sentence. Technically true but does the whole argument hinge on that? Plz explain this to me I'm dumb} \\ 

% \jg{Interpretation: expect high sensitivity away from fixed points? Or something else?}

% 
%    \begin{equation}
% \begin{split}
%     4 \mathcal{W}'  e^{-\mathcal{W}} \partial_z  \left(e^{-i p \cdot x} e^{-3 \mathcal{W}(z) / 2}  \bar{h}_{\mu \nu}(z)\right) = e^{-i p \cdot x}  2 \mathcal{W}' e^{- \mathcal{W}(z) / 2} \bar{h}'_{\mu \nu} - e^{-i p \cdot x}  3 \mathcal{W}' e^{- \mathcal{W}(z) / 2} \bar{h}_{\mu \nu};
% \end{split} 
% \end{equation} 
% 

% 
%    \begin{equation}
% \begin{split}
%       e^{-2\mathcal{W}} \partial^2_z  \left(e^{-i p \cdot x} e^{-3 \mathcal{W}(z) / 2}  \bar{h}_{\mu \nu}(z)\right) = \frac{1}{2}e^{-i p \cdot x} e^{-3 \mathcal{W}(z) / 2} \bar{h}''_{\mu \nu} - e^{-i p \cdot x}  \frac{3}{2} \mathcal{W}' e^{-3 \mathcal{W}(z) / 2} \bar{h}'_{\mu \nu} \\+  e^{-i p \cdot x}  \frac{9}{8} \mathcal{W}'^2 e^{-3 \mathcal{W}(z) / 2} \bar{h}_{\mu \nu} -  e^{-i p \cdot x}  \frac{3}{4} \mathcal{W}'' e^{-3 \mathcal{W}(z) / 2} \bar{h}_{\mu \nu}
% \end{split} 
% \end{equation} 
% 

\section{Three-form solutions\label{sec:sols}}

We now turn to obtaining solutions for the three-form and explicit analysis. The complex nature of our system of equations requires us to use numerical methods. We construct a dynamical system to reduce the derivative order from second to first. We will do so for both the model we have considered thus far, and also with an additional scalar field, acting as a matter source.

% \begin{figure}[h!]
%     \centering
% \includegraphics[width=0.4\textwidth]{fukd.jpeg}
%     \caption{Three-form, potential, and warp factor plotted with exactly determined system. Not symmetric, divergent.}
%     \label{fig:fukd1}
% \end{figure}
\subsection{Solutions without matter}

Using the dynamical variables:
\be
x=\chi\quad z=\chi'+3\mathcal{W}'\chi\quad f=\mathcal{W}'
\ee
our system of equations \eqref{efef} and \eqref{eomf} can be expressed as the following dynamical system:
\begin{eqnarray}
    x'&=&z-3fx\mathcomma\\
    z'&=&-z\left(\frac{z}{x}+f\right)\mathcomma\\
    f'&=&-\frac{4}{3}z^2\mathperiod
\end{eqnarray}

It should be noted that this choice of variables does not impact the form of the dynamical system and thus does not affect the results.\\

\begin{figure}
    \centering
    \includegraphics[width=0.6\linewidth]{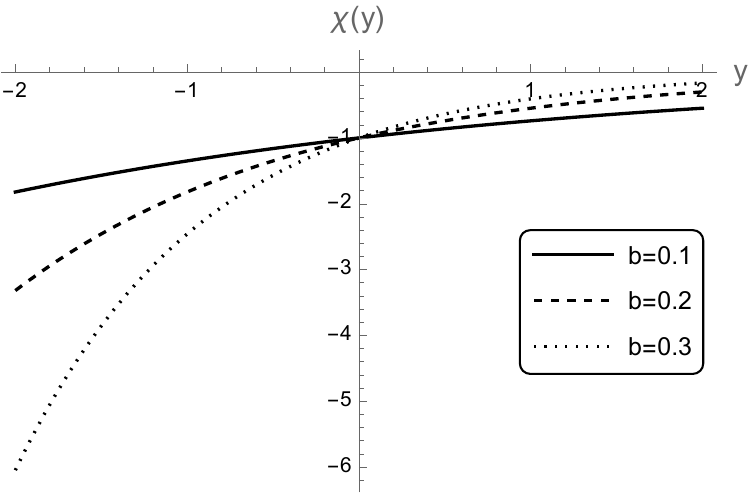}
     \includegraphics[width=0.6\linewidth]{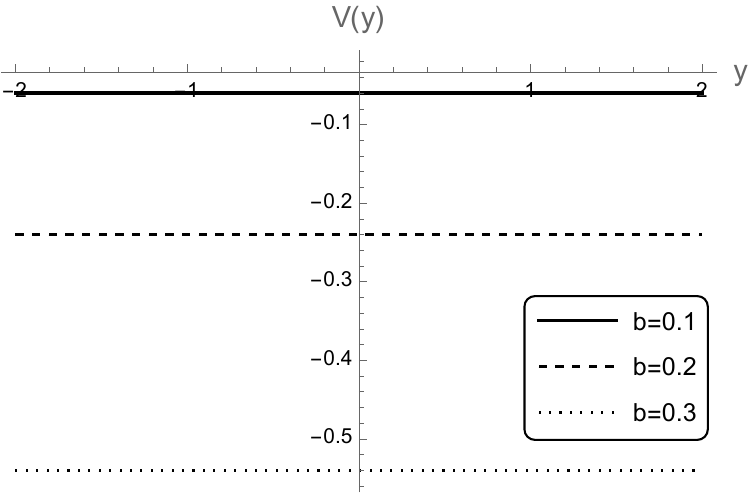}
      \includegraphics[width=0.6\linewidth]{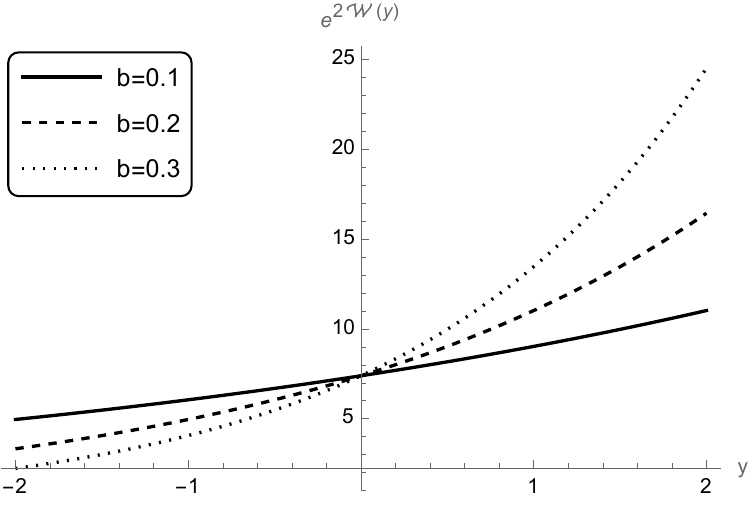}
    \caption{Analytic solutions on the invariant submanifold $z=0$ for the warp factor $e^{2\mathcal{W}(y)}$, the three-form field $\chi (y)$ and its associated potential $V$ for different values of the constant $b$. For all solutions, the values $\mathcal{W}_0= 1$ and $\chi_0 = -1$ are fixed. All the parameters we have fixed are free to fix, and their actual choice has no bearing on the general behaviour.}
    \label{fig:nomatteranalytic}
\end{figure}

We begin the analysis first by considering cases where analytical insight is possible. This system has one line of finite fixed points $\mathcal{P}_L$ at $(x,z,f) = (x, 0, 0)$ corresponding to a 5D Minkowski solution (with no brane). The three-form field is constant in this case, $\chi = \chi_0$\footnote{Here we are not looking at perturbations, $\chi_0$ is not the zeroth order field. Hereafter, subscript $0$ denotes the value of a variable at $y=0$, i.e. $\chi(0) = \chi_0$.}. Additionally, just $z=0$ is an invariant submanifold. On this submanifold, assuming $\mathcal{W}'\neq 0$ and $\chi' \neq 0$, we find solutions which are analytic:
\begin{align}
    \label{eq:analsol}
    \mathcal{W}(y) =& b y + \mathcal{W}_0,\\
    \chi(y) =& \chi_0 e^{-3 b y},\\
    V =& -6b^2,
\end{align}
 where $b$ is a constant. These are plotted in Figure \ref{fig:nomatteranalytic}.  At the level of the action, this submanifold corresponds to $F^2=0$ and a (negative) constant three-form potential in 5D. The means our potential plays the role of an effective cosmological constant and our action is no more than a de Sitter-like action in 5D. \\

% \kelly{used totally arbitrary values for the constant, mostly for illustrative purposes.}

For $z\neq 0$, the system must be solved numerically. To do this, we need to impose boundary conditions. In the case of $f(R,T)$ scalar fields \cite{Rosa:2021tei} boundary conditions at the origin are set to guarantee symmetric solutions, i.e. $\mathcal{W}'(0)=0, \, \chi'(0) = 0$. $Z_2$ symmetry isn't \textit{required} but it is assumed unless there is an explicit reason not to \cite{Dzhunushaliev_2010}. In our case, these conditions correspond directly to the fixed points $\mathcal{P}_L$, admitting only the trivial solution $\chi = \chi_0, \, \mathcal{W} = \mathcal{W}_0,\, V = 0$. \\

Therefore the only case admitting symmetric solutions for both $\chi$ and $\mathcal{W}$ is the Minkowski solution with no brane. It is noted that nontrivial solutions exist which satisfy either condition $\chi'(0)=0$ or $\mathcal{W}'(0)=0$ separately. However, this alone does not produce symmetry -- one would need both conditions to be satisfied. As such, we cannot assume $Z_2$ symmetry for our initial conditions and must solve the system with a different set of conditions.\\

\begin{figure}
    \centering
    \includegraphics[width=0.6\linewidth]{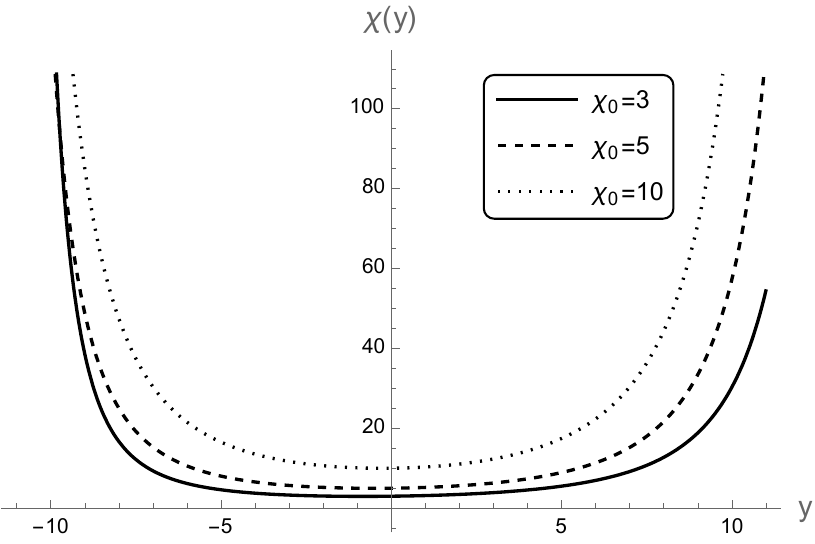}
    \includegraphics[width=0.6\linewidth]{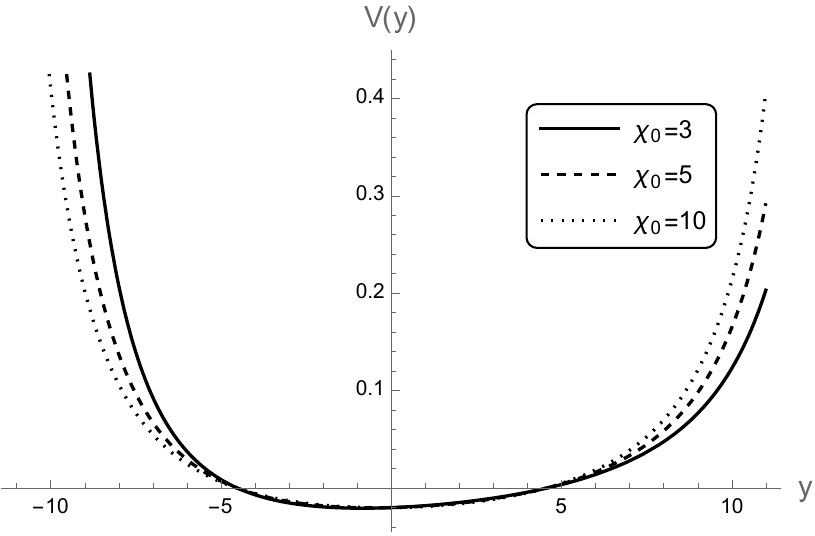}
    \includegraphics[width=0.6\linewidth]{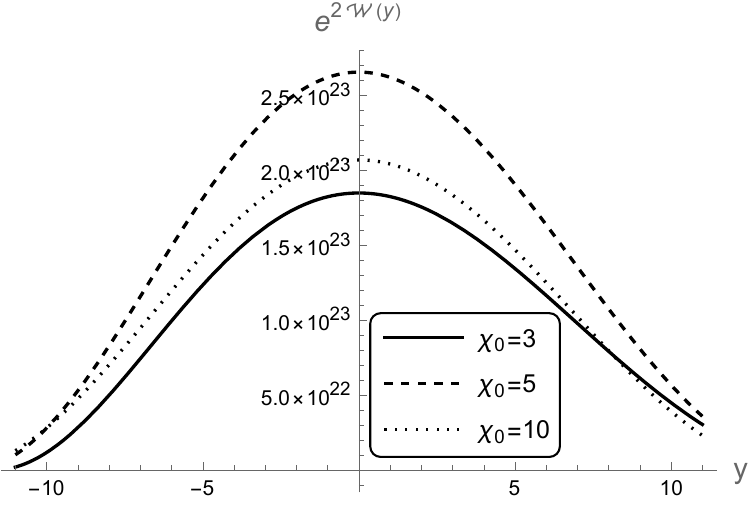}
    \includegraphics[width=0.6\linewidth]{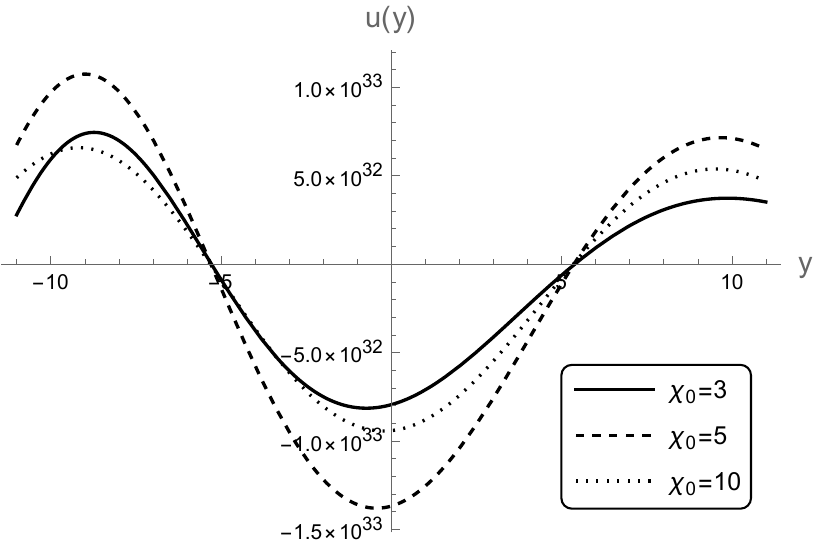}
    \caption{Solutions for different values of $\chi_0$  for the three-form field $\chi(y)$, its associated potential $V(y)$, the warp factor $e^{2\mathcal{W}(y)}$. The bottom panel is the the graviton potential eq. \eqref{gravitoneom} but in terms of $y$, $u(y)$ (see Appendix \ref{app:potentialredef}), with fixed initial conditions corresponding to $\chi'(0)= 0.1$ and $\mathcal{W}'(0)=0$.}
    \label{fig:semisymmetric}
\end{figure}

\begin{figure}
    \centering
    \includegraphics[width=0.6\linewidth]{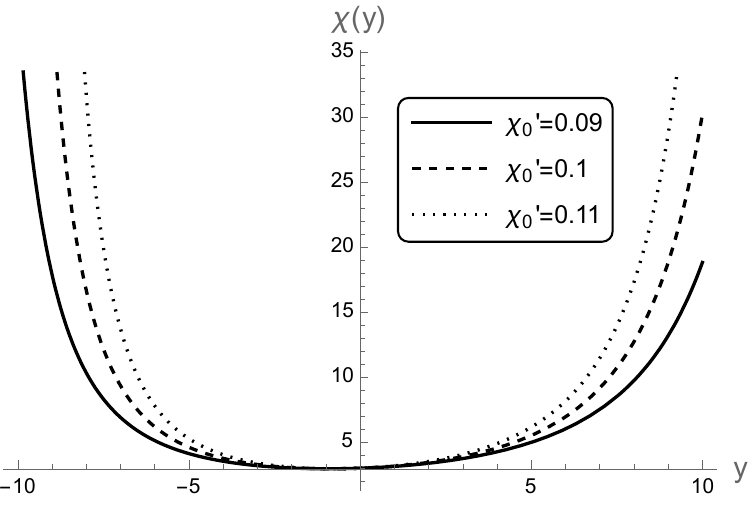}
    \includegraphics[width=0.6\linewidth]{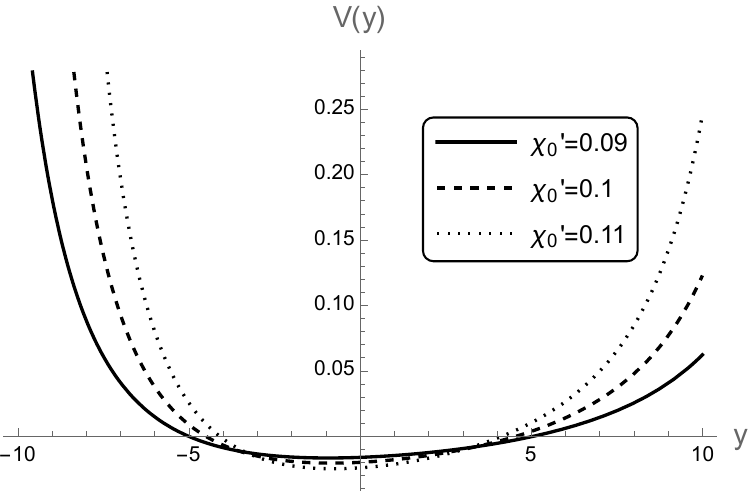}
    \includegraphics[width=0.6\linewidth]{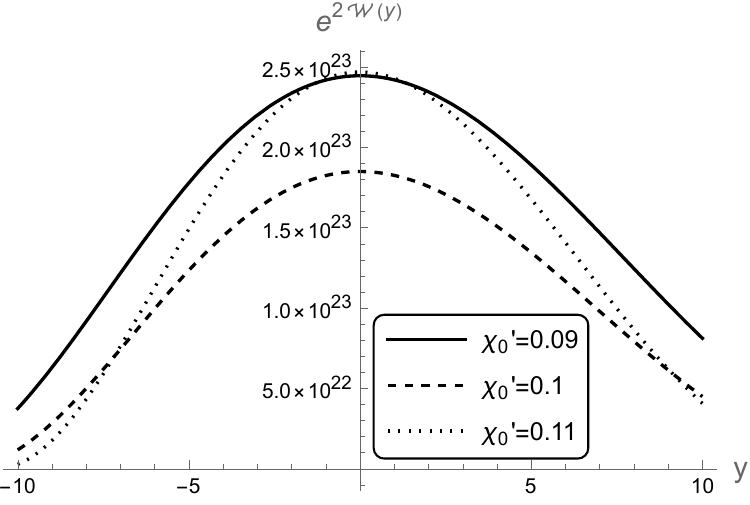}
     \includegraphics[width=0.6\linewidth]{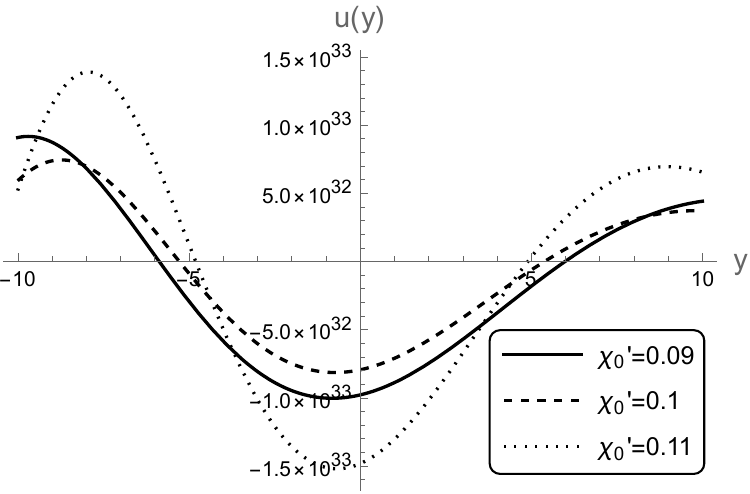}
    
    \caption{Solutions for different values of $\chi_0'$  for the three-form field $\chi(y)$, its associated potential $V(y)$, and the warp factor $e^{2\mathcal{W}(y)}$. The bottom panel is the the graviton potential eq. \eqref{gravitoneom} but in terms of $y$, $u(y)$ (see Appendix \ref{app:potentialredef}), with fixed initial conditions corresponding to $\chi(0)=3$ and $\mathcal{W}'(0)=0$.}
    \label{fig:ff}
\end{figure}

There is no reason to favour one set of initial conditions over another, so for now we shall endeavour to have the solutions be as close to symmetric as possible. We impose only the condition $\mathcal{W}'(0)\equiv \mathcal{W}'_0=0$ and examine solutions for small values of $\chi_0'$ and different values of $\chi_0$. Solutions are shown in Figures \ref{fig:semisymmetric} and \ref{fig:ff}. Figure \ref{fig:semisymmetric} shows the solutions for a range of initial three-form values and Figure \ref{fig:ff} shows the solutions for a range of initial three-form derivative values.\\

The solution for $\chi$ is asymmetric. As mentioned earlier, despite imposing $\mathcal{W}'(0)=0$, this condition is not sufficient to ensure symmetry. The solution for the warp factor will be influenced by the asymmetric nature of $\chi$. This would imply differential behaviour on different sides of the brane, depending on position in the bulk. \\

It is also noted that $\chi$ blows up very rapidly away from the centre of the brane, regardless of the value of either $\chi'_0$ or $\chi_0$. The three-form field exhibits pathological behaviour at some point in its evolution. The precise value of $y$ is initial condition dependent, but the overall result is not. This implies the three-form is unstable off the brane. The instability corresponds to a runaway three-form potential: there are finite minima on the brane, in which the three-form is itself finite, but as the potential grows for large (positive and negative) values of $y$ in the bulk. The warp factor solution obtained is a thick brane solution in the sense of peaking on the brane itself, $y=0$, but it is emphatically not a regular function. \\

This corroborates the perturbation analysis. The energy spectrum is not bounded from below and the background spacetime is unstable against perturbations. This would explain why the field blows up for $y$ values too far from the initial configuration, and why the rest of the behaviour (shape, when it blows up, and so on) is so dependent on initial values. We can see this further by looking at the plot for $u(y)$. The graviton potential has a minima, and despite the apparent trend to flatten out away from the brane, they diverge too (we have kept the $y$ range smaller to keep it consistent with the other plots). This is expected, as we have shown analytically that the metric perturbation does not have a stable energy spectrum.    \\

\subsection{Introducing a matter source}

We now introduce an additional matter source to our model,
\be\label{totalaction}
S=\int d^4xdy\sqrt{-g}\left[ \frac{1}{2\kappa^2}R-\mathcal{L}_{\text{3f}} \right] + S_m(g_{AB},\psi)\mathcomma
\ee
where $S_m$ is the matter action and $\psi$ denotes the matter fields. We will consider matter described by a single  dynamical scalar field $\psi(y)$, dependent only on the extra dimension, with an interaction potential $U(\psi)$. Hence the matter action is
\begin{equation}\label{actionmatter}
    S_m = -\int d^4x dy\sqrt{-g} \left[ \frac{1}{2}\partial^C \psi\partial_C \psi + U(\psi)\right].
\end{equation}

Taking the variation of eq.~\eqref{actionmatter} with respect to the scalar field $\psi$, we obtain the stress-energy tensor for the matter field 
\begin{equation}
    T^{(m)}{}_{AB} = - g_{AB}\left[\frac{1}{2} \partial^C \psi \partial_C \psi + U(\psi)\right] + \partial_A \psi \partial_B \psi.
\end{equation}

The total stress-energy tensor is the sum of the stress-energy tensor for the three-form $T^{(3f)}{}_{AB}$ defined in \eqref{3f_EM} and that of the scalar field $\psi$,
\begin{equation}
    T^{(tot)}{}_{AB} = T^{(3f)}{}_{AB} + T^{(m)}{}_{AB}.
\end{equation}

Varying eq.~\eqref{totalaction} with respect to $\psi$, we obtain the standard Klein-Gordon equation for the matter field,
\begin{equation}\label{kgscalar}
    \psi'' + 4\mathcal{W}'\psi' -U_{\psi} = 0,
\end{equation}
where the prime denotes differentiation with respect to $y$ and the subscript ${}_\psi$ denotes differentiation with respect to $\psi$.\\ 

The full set of field equations now read 
\begin{align}
      3\left(\mathcal{W}''+2\mathcal{W}'^2\right) =& -V-\frac{1}{48}F^2 -\frac{1}{2}\psi'^2 - U\label{efematter1}, \\
    3\left(\mathcal{W}''+2\mathcal{W}'^2\right) =& -V + \frac{1}{48}F^2 +\chi V_{\chi}-\frac{1}{2}\psi'^2 - U,\label{efematter2}\\
    6 \mathcal{W}'^2 =& -V + \frac{1}{48}F^2 + \frac{1}{2}\psi'^2-U.\label{efematter3}
\end{align}

There is freedom to fix the warp function, since between the three EFFs \eqref{efematter1}, \eqref{efematter2}, \eqref{efematter3} and two KG equations \eqref{eom}, \eqref{kgscalar}, we only have four independent equations and five unknowns:

\begin{align}
    &\psi'' + 4\mathcal{W}'\psi' -U_{\psi} = 0,\\
     &\chi ''+ 4 \mathcal{W}' \chi '+3 \chi  \left(\mathcal{W}''+\mathcal{W}'^2\right)-\frac{1}{4}\frac{dV}{d\chi} = 0,\\
      &3\mathcal{W}'' =\chi V_\chi -\psi'^2,\\
     &3\left(\mathcal{W}''+2\mathcal{W}'^2\right) = -V-2(\chi'+3\mathcal{W}'\chi)^2 -\frac{1}{2}\psi'^2 - U. 
\end{align}

With our newfound freedom, we choose the very standard form of the warp factor,  
\begin{equation}
    \mathcal{W}(y) = \mathcal{W}_0 \log[\sech (ky) ],
\end{equation}
where $k, \mathcal{W}_0$ are constants. With this choice, the resulting set of equations are for $\chi$, $V$, $\psi$ and $U$ are

    \begin{align}
 \psi'^2=&3 k^2 \mathcal{W}_0 \sech^2(k y)- 4 (\chi'-3 k \mathcal{W}_0 \chi  \tanh (k y))^2\\
  \chi'' =& 3 k^2 \mathcal{W}_0 \chi  \sech^2(k y) (-2 \mathcal{W}_0 \cosh (2 k y)+2 \mathcal{W}_0+1)+10 k \mathcal{W}_0 \chi' \tanh (k y)-\frac{\chi'^2}{\chi}\\
   V' =& \frac{\chi'}{\chi} (-3k^2 \mathcal{W}_0 \sech^2(k y)+\psi'^2)\\
  U' =& \psi' (\psi'' - 4k\mathcal{W}_0\tanh(ky)\psi' )
\end{align}

%\kelly{I've looked through these equations 100 times and it's driving me mad. I have no idea if this is the right set of independent equations. This set of 4 can obviously be further simplied down and I can get it to the point where the whole system is only in 2 equations in terms of $\psi$ and $\chi$ and their derivatives but that seems weird. Let me know what you think. I'm also a little stressed that fixing the warp factor has made the $\chi$ equation totally independent of all the other variables. This feels weird, like $\chi$s evolution has no dependence on anything else happening, it's totally independent of it's associated potential or anything to do with the matter field. This just feels very weird to me.}

As in the matter-free case, we use dynamical systems methods to solve this by introducing the dynamical variables 

\begin{equation}
\begin{split}
      &x = \chi, \quad z = \chi' -3k\mathcal{W}_0 \tanh(ky)  \chi, \quad \\ 
      &\phi = \psi, \quad \Phi = \psi'.
\end{split}
\end{equation}

The resulting dynamical system is 

\begin{align}
    x' =& z+3k \mathcal{W}_0 \tanh (ky) x\\
    z' =& k\mathcal{W}_0\tanh (ky) z  - \frac{z^2}{2x}\\
    \phi' =& \Phi\\
    \Phi' =& 4k\mathcal{W}_0 \tanh(ky) \Phi +\frac{1}{\Phi}\bigg[\frac{2z^3}{x}+6k\mathcal{W}_0\tanh (ky) z^2  \nonumber\\
    &- 3k^2\mathcal{W}_0 \sech^2(ky)\tanh(ky)(1+2\mathcal{W}_0 )\bigg]  \label{eq:phiev}
\end{align}

% \kelly{There might be a smarter choice for the $\phi$ variable, something like $\Phi = \psi' + 4\mathcal{W}'\psi$ but I'll need time to check if this makes the equations better.  }

\begin{figure}
    \centering
    \includegraphics[width=0.6\linewidth]{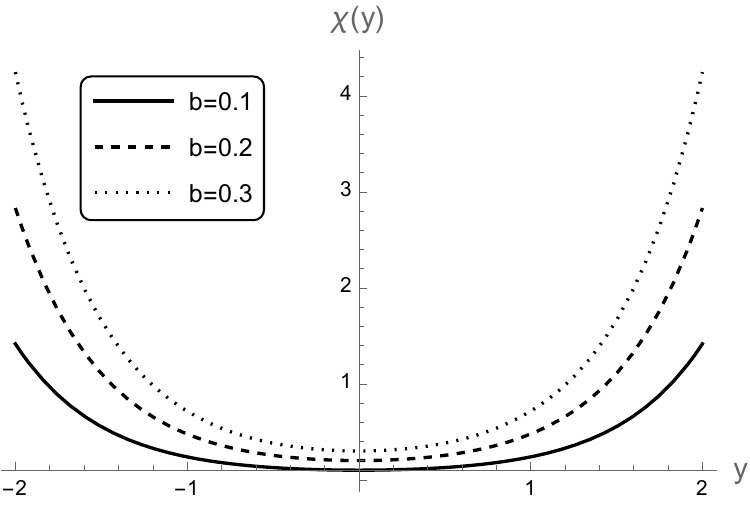}
    \includegraphics[width=0.6\linewidth]{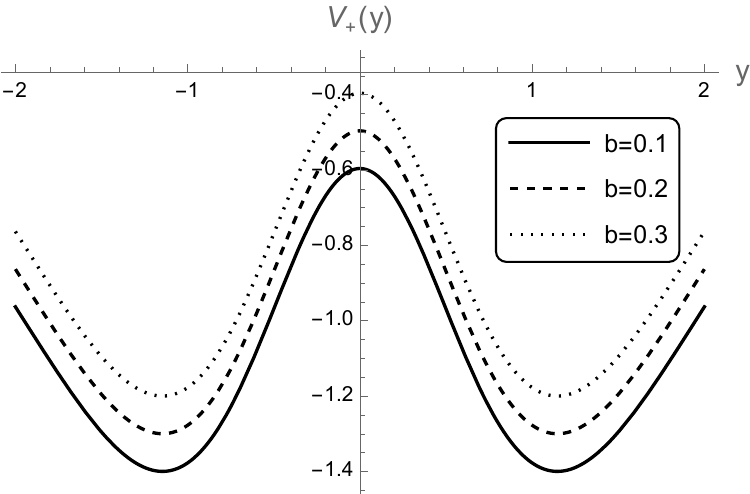}
    \includegraphics[width=0.6\linewidth]{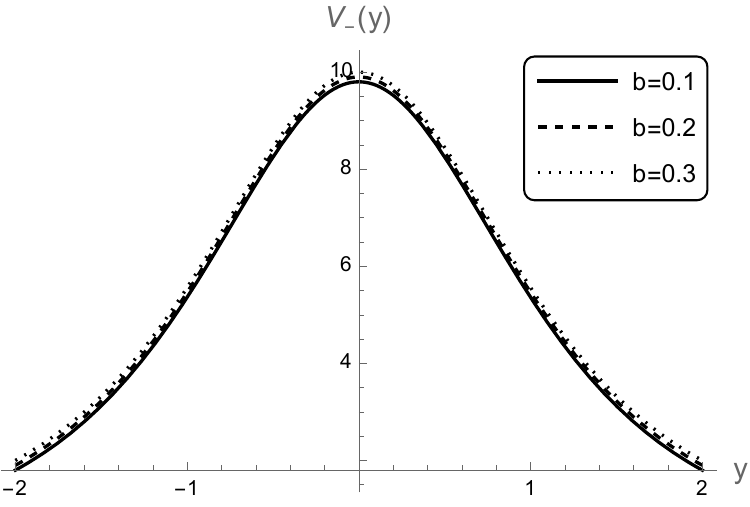}
    \caption{Analytic solutions for the matter case on the invariant submanifold $z=0$ for the three-form field $\chi (y)$ and its associated potentials $V_+$ and $V_-$ for different values of the constant $b$. For all solutions, the values $\mathcal{W}_0= 1$ and $k = 1$ are fixed.}
    \label{fig:mattanalytic1}
\end{figure}

\begin{figure}
    \centering
    \includegraphics[width=0.6\linewidth]{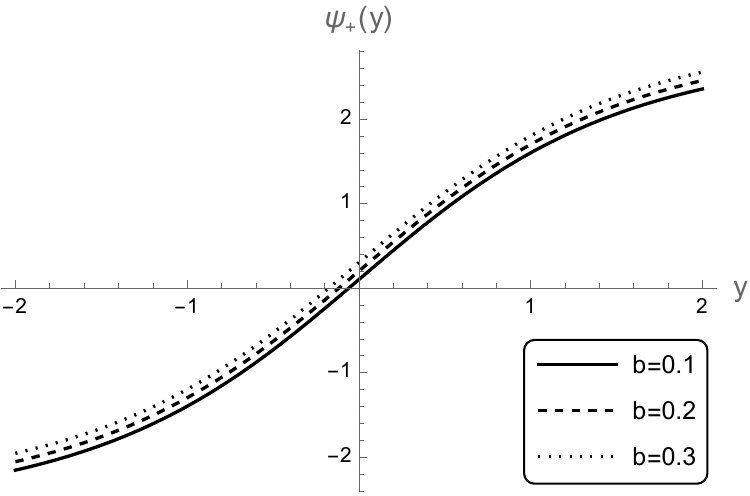}
    \includegraphics[width=0.6\linewidth]{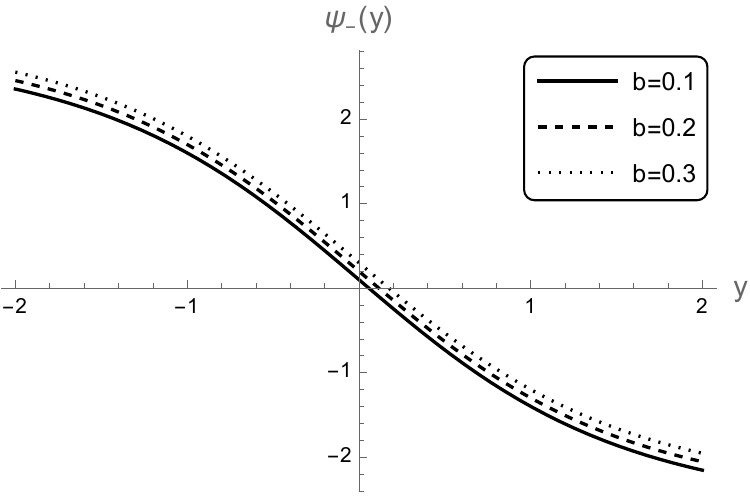}
    \includegraphics[width=0.6\linewidth]{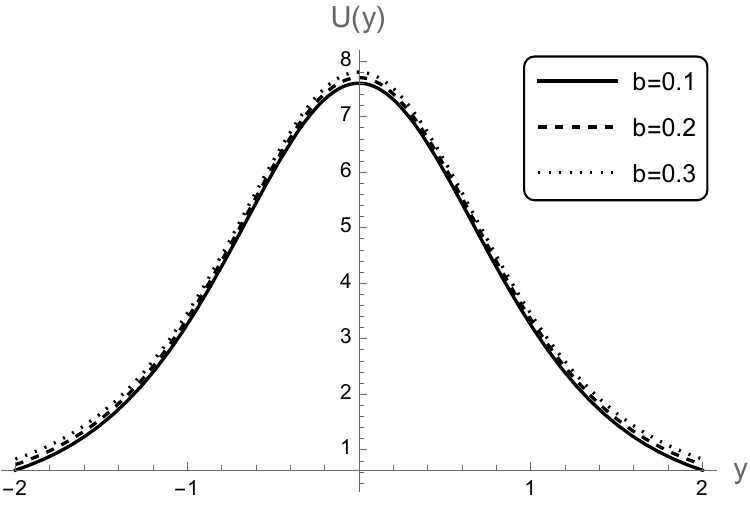}
    \caption{Analytic solutions for the matter case on the invariant submanifold $z=0$ for the matter field $\psi_+ (y)$ and $\psi_- (y)$ and the associated potential $U(y)$ for different values of the constant $b$. For all solutions, the values $\mathcal{W}_0= 1$ and $k = 1$ are fixed.}
    \label{fig:mattanalytic2}
\end{figure}

This system has many features in common with the matter-free case. $z=0$ is an invariant submanifold on which an analytical solution can be found:

    \begin{align}
    \chi(y) =& b \cosh ^{3 \mathcal{W}_0}(k y)\\
    \psi(y) =& b \pm \frac{2 \sqrt{6} \sqrt{\mathcal{W}_0} \cosh (k y) \tan ^{-1}\left(\tanh \left(\frac{k y}{2}\right)\right)}{\sqrt{\cosh (2 k y)+1}}\\
    V(y) =& b-3 k^2 \mathcal{W}_0^{3/2} \left(\pm\frac{\sqrt{3}}{k \sqrt{\cosh ^2(k y)}}-\frac{3}{2} \sqrt{\mathcal{W}_0} \sech^2(k y)\right)\\
    U(y) =& b + 6 k^2 \mathcal{W}_0^2 \sech^2(k y)+\frac{3}{2} k^2 \mathcal{W}_0 \sech^2(k y),
\end{align}

where $\mathcal{W}_0$, $k$ and $b$ are constants. These are plotted in Figures \ref{fig:mattanalytic1} and \ref{fig:mattanalytic2}. The inclusion of matter has fundamentally changed the dynamics. The potential now evolves, and we can see that the negative solution $V_-$ should be discounted as ``unphysical'', since its behaviour does not correspond with the three-form's. The positive solution does, however, and too exhibits instability at larger values of $y$. The matter fields inherit the apparent instability now that they're dependent on the three-form. Despite the matter potential being well-behaved, the matter field dynamics are not.  \\

\begin{figure}[h!]
    \centering
    \includegraphics[width=0.6\linewidth]{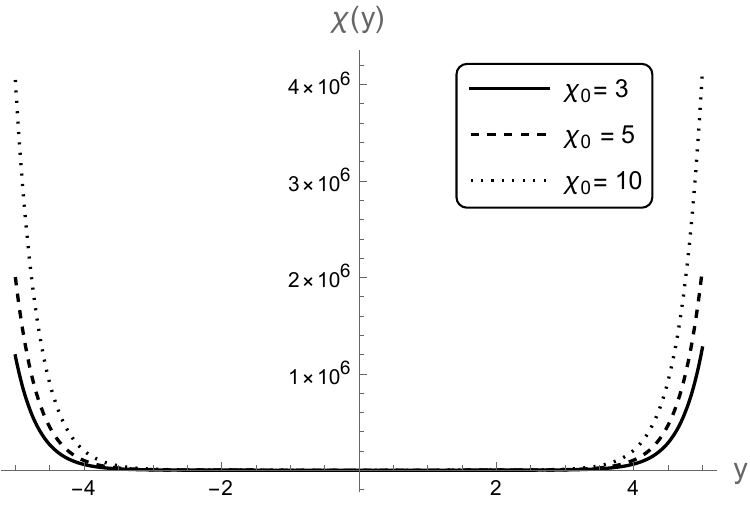}
    \includegraphics[width=0.6\linewidth]{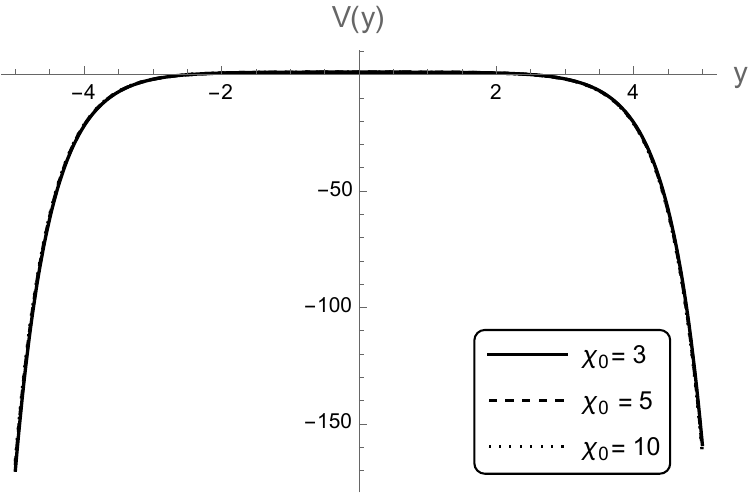}
    \includegraphics[width=0.6\linewidth]{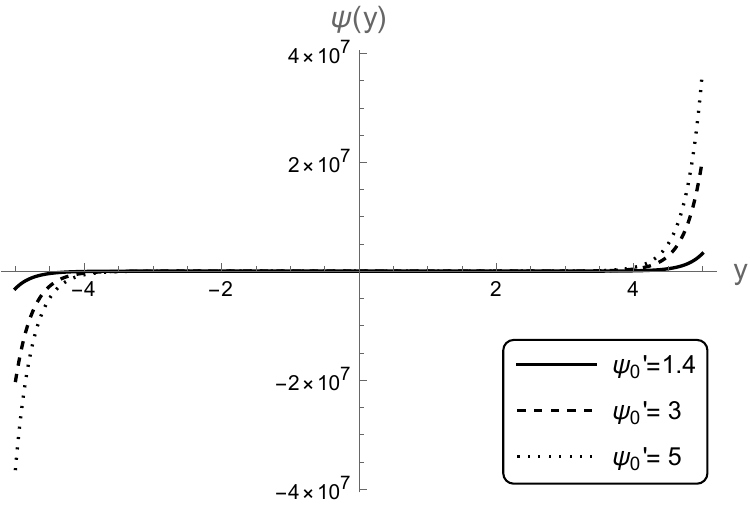}
    \includegraphics[width=0.6\linewidth]{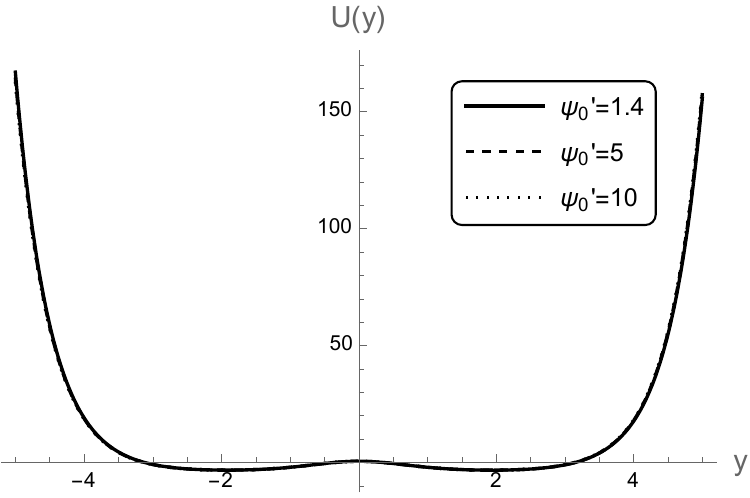}
    \caption{Solutions for different values of $\chi_0$ and $\phi_0$ for the three-form field $\chi(y)$, its associated potential $V(y)$, the matter field $\psi(y)$ and its associated potential $U(y)$ with fixed conditions corresponding to $\chi'(0)=0.1$ and $\psi(0) = 1$. }
    \label{fig:withmatt0}
\end{figure}

\begin{figure}
    \centering
    \includegraphics[width=0.6\linewidth]{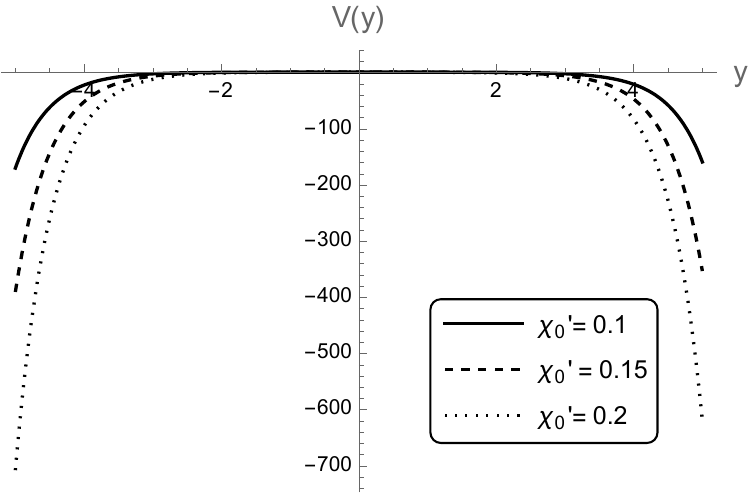}
    \includegraphics[width=0.6\linewidth]{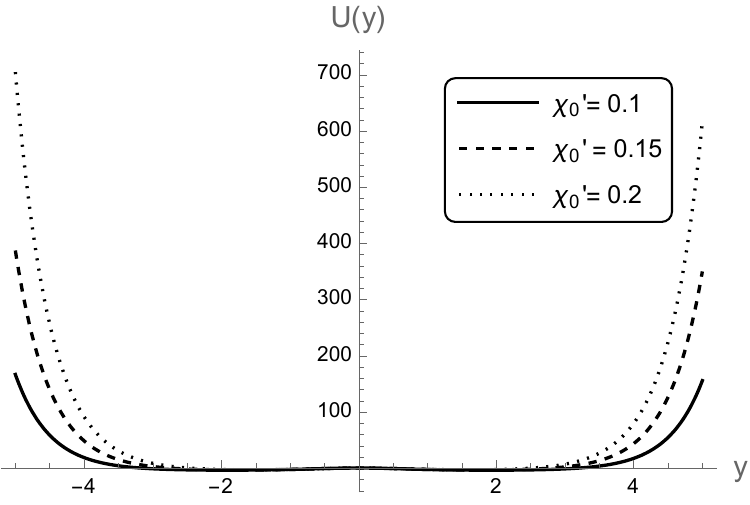}
    \caption{Solutions for different values of $\chi_0'$ for the three-form potential $V(y)$ and the matter potential $U(y)$ with fixed conditions corresponding to $\chi(0)= 3$, $\psi(0) = 1$ and $\psi'(0)=3$.}
    \label{fig:withmatt1}
\end{figure}

Assuming $z\neq 0$, solutions must be found numerically. Again, $z\neq 0$ implies that $\chi'(0) \neq 0$ so numerical, symmetric solutions for $\chi$ are not possible. Additionally, it is not possible %\kelly{at least with this choice of variables},
to get symmetric solutions for $\psi$ since eq.~\eqref{eq:phiev} is singular at $\Phi = 0$. Therefore, similar to the case for $\psi = 0$, initial conditions are imposed on $\chi(0)=\chi_0$ and $\chi'(0)=\chi_0'$ to achieve near-symmetric solutions which remain finite near the brane. As is the case for $\chi_0$ and $\chi_0'$, the system is highly sensitive to values of $\phi_0$ and $\phi_0'$. Solutions for a range of $\chi_0$ and $\phi_0$ values are shown in Figure 
\ref{fig:withmatt0} and we plot the two potentials for a range of $\chi_0'$ values; these are shown in Figure \ref{fig:withmatt1}. \\

Much like the analytical sub-case, the inclusion of matter and an additional degree of freedom does not change the fact that the three-form field  blows up at sufficiently high values off the middle of the brane. It does, however, keep the three-form stable for a larger range of $y$. The three-form potential is not able to maintain symmetry when $\chi_0'$ is too high. For smaller values, as seen in Figure \ref{fig:withmatt1}, the three-form potential maintains some degree of symmetry and does not blow up as quickly compared to the matter-free. However, in general, for both, the initial `kick' must not be too large -- further evidence that three-forms are unstable. Our brane now is symmetric by construction, and yet the instabilities in the dynamics of the three-form manifest themselves nonetheless\\

The scalar field, acting as matter, produces a more stable potential and dynamical behaviour (Figure \ref{fig:withmatt1}. But since the dynamics are connected to the three-form, they blow up too.

% \kelly{It might be a little more clear to see the feedback loop in $z$ and $x$ (Previously from (4.8)-(4.12) ) in this case. We can express the $z'$ equation as 
% \begin{align}
%      z'
%     =& -\frac{z}{2}\left[5W'- \frac{x' }{x}\right]\\
%     =& -\frac{z}{2}\left[8W'- \frac{z }{x}\right]
% \end{align}
% So if $\frac{x'}{x}$ doesn't stay finite, this kicks off the feedback loop which drives $z$ to diverge and thus $\chi$ to diverge, which seems to always been the case. If $z$ and $x$ balance each other (which seems to be the case if $z$ is really small) then you can get finite solutions over a certain range of $y$. However, $z$ will eventually grow and kick off the feedback loop, making solutions diverge at some $y$ value.}

\section{Discussions and Conclusions\label{sec:discussion}}

In this work, we have studied thick
braneworld solutions in general relativity, in a 5D bulk, with the inclusion of a three-form field. This has extended the work done by \cite{Barros:2015evi}, who considered three-form inflation in Randall-Sundrum II braneworlds. We have constructed novel solutions for the three-form field in the extra dimension in a warped Minkowski spacetime. We first developed the general formalism required to investigate the problem, and found that we could not study specific solutions to the problem without matter, due to the exactly determined set of governing equations. We have also studied three-forms in branes with matter, expanding our model to include a scalar field as a matter source. This allowed us to fix the warp function specifically and study that particular configuration.\\

Section \ref{sec:pert} contains new results of interest. The three-form radically changes the study of the stability of the gravity sector of the model. Following the general procedure \cite{DeWolfe_2000}, we have derived the stability differential equation for the graviton. The potential, normally a stability potential for scalar-tensor theories, is not stable in the three-form case. The gravity zero mode is therefore not normalisable. This indicates the background is not stable against perturbations, and that the brane cannot support internal structure (i.e. thick branes specifically are unstable).\\

This finding is further supported by the new results in Section \ref{sec:sols}. We have found new numerical solutions for the three-form field, its potential, and the warp function. We can see from the solutions that the three-form is unstable, diverging off the center of the brane (the center being the initial starting point for our evolution). As expected from Section \ref{sec:pert}, the system is highly sensitive to initial conditions. The system also enjoys no degree of symmetry, displaying differential behaviour for the same point but on the other side of the brane, further enforcing the notion that the dynamics are unstable. Adding in matter does not remedy this situation. Despite affording the freedom to fix a symmetric and well-behaved warp function, the system's evolution still diverges and shows a varying degree of symmetry in the three-form and the scalar, depending on the conditions. In either case, however, there is no means to obtain regular behaviour past a certain point in the system's evolution. \\

There is substantial caveat to the analysis performed in this paper. The one true free choice available is the freedom pick eq. \eqref{dualansatz}. This ansantz is arbitrary, and greatly affects both the three-form equation of motion and the stress-energy tensor in the EFEs. If the three-form were to be parameterised differently, the system of equations would be different -- possibly well-behaved. The perturbations would potentially be stable too. What this paper represents then is a first step in studying three-forms in thick branes, in conjunction with their scalar field counterparts. The indication from this work is that there exist properties of thick braneworlds that render three-forms unstable. A comprehensive proof of this -- by showing this to be the case for a general three-form parameterisation -- is still needed, as is a more thorough theoretical analysis of the properties of thick branes which render them incompatible with three-forms. This remains to be done as future work.

\acknowledgments
The authors thank Bruno Barros for the genesis of the idea, and in particular for discussions relating to the formalism in Section \ref{sec:model} and \ref{sec:pert}. We additionally thank him for help at the blackboard and for useful \textbf{Mathematica} notebooks. We also thank Peter Dunsby for discussions about the implications of this work. KM thanks the University of Cape Town for funding assistance.  

\clearpage

\newpage

\bibliographystyle{apsrev4-1}
\bibliography{bib.bib}

\appendix
\section{Stability analysis through localisation}

There is an alternative way to study stability of tensor forms in branes that will further lend support to our conclusions. We employ the methods of \cite{zhong2023localization}: in this work they use the conformal coupling of a scalar to localise a vector field on the brane. However, their method is applicable to any $q$-form and for any sort of matter coupling, including minimal coupling. \\

We can glean some insight from their condition for stability, which we quote here. The general condition is, for a minimally coupled q-form,

\be
\int^{+\infty}_{-\infty}\left (e^{2\mathcal{W}(y)}\right)^{3-2q} < \infty
\ee \medskip

In our case, we can't fix any warp factor; ours is determined solely by the three-form parameterisation choice. In any event, even without some analytical expression to use here, we can numerically use the result for $\mathcal{W}(y)$ in Figure \ref{fig:semisymmetric} and find the integral here does, of course, diverge. Visually, our warp factor diverges after certain values of $y$, which means the integral above won't converge everywhere. Hence, we can conclude once more, via this method, that three-forms aren't stable in thick branes.

\section{Scalar ``mimicing''?}

Is the under-determined set of equations the missing ingredient to finding consistent and stable solutions? To answer this, a very naive approach is to try `force' the three-form equations to look like their scalar field counterparts (cf. \eqref{scalareomf}). This means we want the components of the $T_i^i$ to match $T_0^0$ in eq.~\eqref{efef}, which means $2T^0_0 =- 2V + \chi V_\chi$, and so 

\be
\begin{split}
    \chi ''+ 4 \mathcal{W}' \chi '+3 \chi  \left(\mathcal{W}''+\mathcal{W}'^2\right)-\frac{1}{4}\frac{dV}{d\chi} &= 0\\
    3\left(\mathcal{W}''+2\mathcal{W}'^2\right) &= -V-\frac{1}{48}F^2\\    
    6 \mathcal{W}'^2 &= \frac{1}{48}F^2 - V 
\end{split}
\ee

Eliminating the $V$ term, we have 

\be
\begin{split}
     \chi ''+ 4 \mathcal{W}' \chi '+3 \chi  \left(\mathcal{W}''+\mathcal{W}'^2\right)-\frac{1}{4}\frac{dV}{d\chi} &= 0\\
    3\left(\mathcal{W}''+4\mathcal{W}'^2\right) &= - 4\left( \chi'+3\mathcal{W}'\chi \right)^2
\end{split}
\ee

We use the chain rule an to rewrite the potential derivative, $dV/d\chi = dV/dy dy/d\chi = V'/\chi'$. We therefore have 

\be
\begin{split}
     \chi ''+ 4 \mathcal{W}' \chi '+3 \chi  \left(\mathcal{W}''+\mathcal{W}'^2\right)-\frac{1}{4} \frac{V'}{\chi'} &= 0\\
    3\left(\mathcal{W}''+4\mathcal{W}'^2\right) &= - 4\left( \chi'+3\mathcal{W}'\chi \right)^2
\end{split}
\ee

In principle we should be able to pick one unknown; for example, we can pick $\mathcal{W} = \log(\sech y)$ like \cite{Rosa:2021tei} and solve. However, due to the `forcing' we have a constraint equation that can be written as 

\begin{equation}
   4\left( \chi'+3\mathcal{W}'\chi \right)^2 + \frac{\chi}{\chi'} V' = 0.
\end{equation}

This essentially exactly determines our system again. Indeed, picking $\mathcal{W} = \log(\sech y)$ solves some of the equations here but not all. This alone is a sufficient counterexample to illustrate that there's no simple forcing the system to be under-determined. 

\section{Graviton potential redefinition \label{app:potentialredef}}

Our graviton potential plots in Figures \ref{fig:semisymmetric} and \ref{fig:ff} plot them as a function of $y$. We quote them as a function of $z$ in eq.~\eqref{gravitoneom}. The potential in terms of $y$ is

\begin{equation}
    \begin{split}
        u(y) = \frac{e^{2\mathcal{W}(y)}}{4}\left[\mathcal{W}'(y)^2\left(9e^{\mathcal{W}(y)}-\frac{21}{4}\right)\right. \left.+\mathcal{W}''(y)\left(3e^{\mathcal{W}(y)}-3\right)\right] 
    - e^{\mathcal{W}(y)}V
 \end{split}
\end{equation}\\

For prosterity, we quote here the full backwards-transformation from $W(z)\rightarrow \mathcal{W}(y)$:

\begin{eqnarray}
    W(z) = \frac{1}{2}\left(-\ln(4/3) + \mathcal{W}(y)\right)\\
    W'(z) = \frac{1}{2} e^{\mathcal{W}(y)} {\mathcal{W}'(y)}\\
     W''(z) = \frac{1}{2} e^{2\mathcal{W}(y)} \left(  {\mathcal{W}'(y)}^2 + {\mathcal{W}''(y)}\right)    
\end{eqnarray}\\

where a prime denotes differentiation w.r.t. the function argument.

\end{document}